\documentclass[conference]{IEEEtran}
\IEEEoverridecommandlockouts
\usepackage{cite}
\usepackage{amsmath,amssymb,amsfonts}
\usepackage{algorithmic}
\usepackage{graphicx}
\usepackage{textcomp}
\usepackage{xcolor}
\usepackage{hyperref}
\def\BibTeX{{\rm B\kern-.05em{\sc i\kern-.025em b}\kern-.08em
    T\kern-.1667em\lower.7ex\hbox{E}\kern-.125emX}}
\begin{document}

\title{Quantum Image Processing}

\makeatletter
\newcommand{\linebreakand}{%
  \end{@IEEEauthorhalign}
  \hfill\mbox{}\par
  \mbox{}\hfill\begin{@IEEEauthorhalign}
}
\makeatother

\author{\IEEEauthorblockN{Alok Anand}
\IEEEauthorblockA{\textit{Electrical \& Computer Engineering} \\
\textit{Carnegie Mellon University}\\
Pittsburgh, PA, U.S.A. 15213 \\
\href{mailto:aloka@andrew.cmu.edu}{aloka@andrew.cmu.edu}}\\
\and
\IEEEauthorblockN{Meizhong Lyu}
\IEEEauthorblockA{\textit{Material Sciences and Engineering} \\
\textit{Carnegie Mellon University}\\
Pittsburgh, PA, U.S.A. 15213 \\
\href{mailto:meizhonl@andrew.cmu.edu}{meizhonl@andrew.cmu.edu}}\\
\and
\IEEEauthorblockN{Prabh Simran Baweja}
\IEEEauthorblockA{\textit{Electrical \& Computer Engineering} \\
\textit{Carnegie Mellon University}\\
Pittsburgh, PA, U.S.A. 15213 \\
\href{mailto:pbaweja@andrew.cmu.edu}{pbaweja@andrew.cmu.edu}}\\
\linebreakand

\IEEEauthorblockN{Vinay Patil}
\IEEEauthorblockA{\textit{Electrical \& Computer Engineering} \\
\textit{Carnegie Mellon University}\\
Pittsburgh, PA, U.S.A. 15213 \\
\href{mailto:vspatil@andrew.cmu.edu}{vspatil@andrew.cmu.edu}}
    
}

\maketitle

\begin{abstract}
Image processing is popular in our daily life because of the need to extract essential information from our 3D world, including a variety of applications in widely separated fields like bio-medicine, economics, entertainment, and industry. The nature of visual information, algorithm complexity, and the representation of 3D scenes in 2D spaces are all popular research topics.  In particular, the rapidly increasing volume of image data as well as increasingly challenging computational tasks have become important driving forces for further improving the efficiency of image processing and analysis.
Since the concept of quantum computing was proposed by Feynman in 1982, many achievements have shown that quantum computing has dramatically improved computational efficiency\cite{ruan2021quantum}. Quantum information processing exploit quantum mechanical properties, such as quantum superposition, entanglement and parallelism, and effectively accelerate many classical problems like factoring large numbers, searching an unsorted database, Boson sampling, quantum simulation, solving linear systems of equations, and machine learning. These unique quantum properties may also be used to speed up signal and data processing. In quantum image processing, quantum image representation plays a key role, which substantively determines the kinds of processing tasks and how well they can be performed. 
\end{abstract}

\vspace{1em}
\begin{IEEEkeywords}
Feature Map, QPIE, Hadamard, QSobel, Quantum CNN, Edge Detection, K-means, Corner detection, Quantum kernel, SVM, CIFAR, Noise Probability, Pauli Noise, Depolarising noise
\end{IEEEkeywords}

\section{Literature Review}
\subsection{Classical Image Representation}
Images in classical computers are defined as matrices of numbers representing the discrete color or intensity values present in every image pixel. Each image is considered as input data displayable in numerous ways, whether as arrays of pixel values or either multidimensional plots representing the distribution of pixel intensities. Images can be rendered in color layered with 3 channels (Blue, Green, and Red), Grayscale with pixel values varying from 0 (black) to 255 (white), and binary portraying black or white values (0 or 1) only. 

\subsection{Quantum Image Representation (QIR)}
Quantum image processing is one of the most attractive and promising tools within the Quantum Technology toolbox. The representation of an image on a quantum computer in the form of normalized states facilitates numerous image processing problems\cite{yan2016survey}. 

\begin{figure}[h!]
    \begin{center}
    \includegraphics[width=0.5\textwidth]{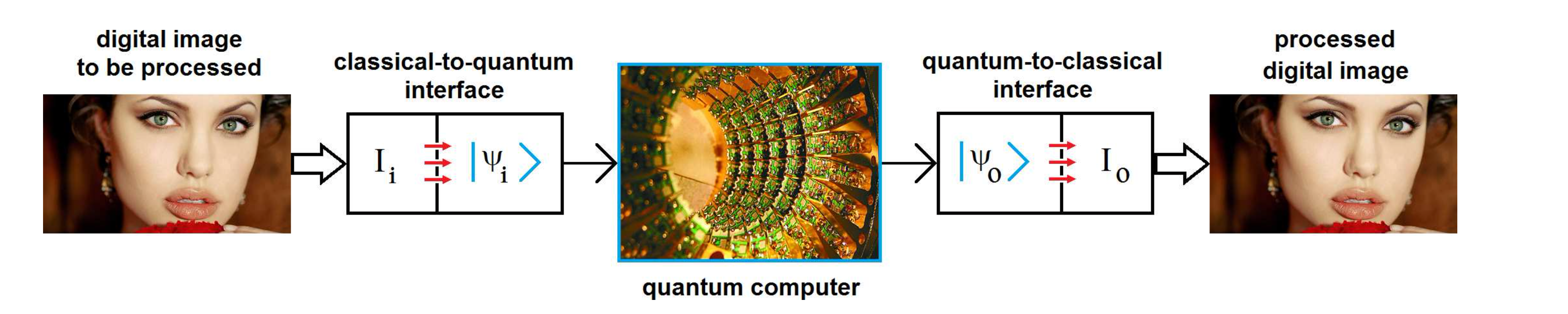}
    \end{center}
    \caption{Scheme of Quantum Image Processing (QImP)}
    \label{fig:QImp}
\end{figure}

The following methods have been proposed for QIR: 

\subsubsection{\textbf{Qubit Lattice}}
In 2003, Venegas-Andraca and Bose suggested the \textit{Qubit Lattice} model \cite{10.1117/12.485960} to map image’s spatial information with the amplitude of a single qubit without using quantum properties, therefore requiring the same number of qubits as pixels. This is a quantum-analog representation of classical images. The pixel value of $i^{th}$ row and the $j^{th}$ column can be stored as:
\begin{equation}
|pixel_{i,j}\rangle=(\cos \left(\frac{\theta_{i,j}}{2}\right)|0\rangle+\sin \left(\frac{\theta_{i,j}}{2}\right)|1\rangle)
\end{equation}

\subsubsection{\textbf{Flexible Representation of Quantum Images (FRQI)}}
FRQI, proposed by Le et al.\cite{le2011flexible}, maps each pixel’s grayscale value to the amplitude as well as captures the corresponding positions in an image and integrates them into a quantum state. The FRQI representation is expressed as:

\begin{equation}
|I(\theta)\rangle=\frac{1}{2^{n}} \sum_{i=0}^{2 n-1}\left(\sin \left(\theta_{i}\right)|0\rangle+\cos \left(\theta_{i}\right)|1\rangle\right)|i\rangle
\end{equation}

where $\theta_i$ encodes pixel value of the corresponding position $|i⟩$.
The FRQI representation maintains a normalized state and the representation space decreases exponentially compared to the classical image due to the quantum states' superposition effect. FRQCI and IFRQI are two improved flexible representation of quantum images reported by Li et al.\cite{li2016quantum} and Khan et al.\cite{khan2019improved} respectively.

\subsubsection{\textbf{Novel Enhanced Quantum Representation (NEQR)}}
Zhang et al.\cite{zhang2013neqr} reported a representation that uses the basis state of a qubit sequence to store the grayscale value of every pixel instead of probability amplitude encoded in a qubit as in FRQI. The images are stored by entangling color information, represented by $|f(y, x)\rangle$, with location information, represented by $|y x\rangle$.  The NEQR representation for a ${2^{n}} \times {2^{n}}$ image is expressed as:

\begin{equation}
|I\rangle=\frac{1}{2^{n}} \sum_{y=0}^{2^{2 n}-1} \sum_{x=0}^{2^{2 n}-1}|f(y, x)\rangle|y x\rangle
\end{equation}

where $f(y,x)$ refers to the pixel intensity at $f(y,x)$. 
NEQR could perform the complex and elaborate color operations more conveniently than FRQI does. NEQR can achieve a quadratic speedup in quantum image preparation and retrieve digital images from quantum images accurately. However, NEQR representation uses more qubits to encode a quantum image. Researchers have proposed a few variations for NEQR as well: Improved NEQR (INEQR) \cite{zhou2018quantum}, generalized model of NEQR (GNEQR) \cite{li2019block} and CQIR \cite{caraiman2013image}.

\subsubsection{\textbf{Normal Arbitrary Quantum Superposition State (NAQSS)}}
NAQSS was proposed to solve multi-dimensional color image processing by Li et al.\cite{li2014multi}. The NAQSS is $a (n + 1)$- qubit quantum representation, which can be used to represent multi-dimensional color images. N qubits represent colors and coordinates of $2^{n}$ pixels and the remaining 1 qubit represents an image segmentation information to improve
the accuracy of image segmentation. The NAQSS corresponds to the color and angle of the image one to one, mapping the color information to a certain value on the interval $[0, \pi/2]$. The NAQSS can be represented as:

\begin{equation}
|I\rangle=\sum_{i=0}^{2^{n}-1}\theta_{i}|v_{1}\rangle|v_{2}\rangle…|v_{k}\rangle\otimes|\chi_i\rangle
\end{equation}
where
\begin{equation}
|\chi_i\rangle=\cos\gamma_{i}|0\rangle + \sin\gamma_{i}|1\rangle
\end{equation}
represents the segmentation information. NAQSS can improve the efficiency and accuracy of image segmentation. However, it cannot accurately measure the pixels of the image.

\subsubsection{\textbf{Quantum Probability Image Encoding (QPIE)}}
Yao et al.’s QPIE \cite{yao2017quantum} representation has an immediate advantage over the other encoding methods previously discussed is that it allows to encode rectangular r × c images. Moreover, the number of qubits necessary to encode the image are further reduced. The QPIE representation is expressed as:
\begin{equation}
|I\rangle=\sum_{i=0}^{2^{2 n}-1} c_{i}|i\rangle, n=\left[\log _{2}(r c)\right]
\end{equation}
where
\begin{equation}
 I^{\prime}=\left(I_{1,1}, I_{2,1}, \ldots, I_{r, 1}, I_{1,2}, \ldots, I_{r, c}\right)^{T}
\end{equation}
and 
\begin{equation} c_{i}=\frac{I^{\prime}(i)}{||I^{\prime}||}\end{equation} is the normalized value of the ${i^{th}}$ element of the vector.

This method allows one to encode images of arbitrary size r × c  However, this encoding has the problem of the extraction of the exact original image from its encoding quantum circuit. This is because the pixel value is stored in the state amplitude, so that its approximate value can only be restored by multiple measurements.

\subsubsection{\textbf{Quantum Feature Map based Data Encoding}}
Classical data is encoded to the quantum state space using a quantum feature map. The choice of the feature map to use is important and depends on the dataset to be classified. A quantum feature map utilises classical feature vector for encoding to quantum state, that includes applying the unitary operation on the initial state.
\begin{figure}[h!]
  \begin{center}
  \includegraphics[width=9cm]{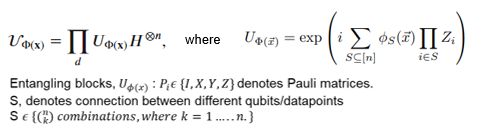}
  \end{center}
  \caption{Feature map on N-qubits generated
by the unitary function from Havlicek et al.’s work\cite{2019}}
  \label{fig:a Feature map on N-qubits generated
by the unitary function}
\end{figure}

Feature map contain layers of Hadamard gates with entangling blocks encoded as,

\begin{figure}[h!]
  \begin{center}
  \includegraphics[width=9cm]{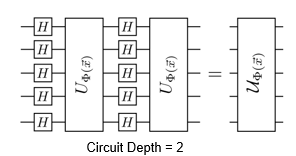}
  \end{center}
  \caption{Feature map on N-qubits generated
by the unitary function from Havlicek et al.’s work\cite{2019}}
  \label{fig:a Feature map on N-qubits generated
by the unitary function}
\end{figure}
Here, data encoding employs ZZFeature Map , Pauli feature and Zfeature map based quantum SVM kernel, as introduced in Havlicek et al \cite{2019}.
In preparation of quantum kernel matrices for training and testing, feature map is first applied to each pair of training datapoints, then to testing datapoints. In later stage, train and test quantum kernel matrices are utilised for classification support vector machine. 

\subsection{Classical Edge and Corner Detection}
\subsubsection{Edge Detection}
Edge detection is an image processing technique for finding the boundaries of an object in the given image. The edges are the part of the image that represents the boundary or the shape of the object in the image. the general method of edge detection is to study the changes of a single image pixel in a gray area, use the variation of the edge neighboring first-order or second-order to detect the edge. There are various methods in edge detection, and following are some of the most commonly used methods:

\begin{enumerate}
    \item \textbf{Prewitt edge detection \cite{canny1983finding}}: This method is commonly used to detect edges based applying a horizontal and vertical filter in sequence. This gradient based edge detector is estimated in the 3x3 neighborhood for eight directions. All the eight convolution masks are calculated.
    
    \item \textbf{Sobel edge detection \cite{irwin1968isotropic}}: Sobel detection finds edges using the Sobel approximation to the derivative. It precedes the edges at those points where the gradient is highest. It is one of the most commonly used edge detectors and helps reduce noise and provides differentiating, giving edge response simultaneously.
    
    \item \textbf{Laplacian edge detection \cite{peli1982study}}: This method uses only one filter (also called a kernel). In a single pass, Laplacian edge detection performs second-order derivatives and hence are sensitive to noise. The Laplacian is generally used to found whether a pixel is on the dark or light side of an edge.

    \item \textbf{Canny edge detection \cite{canny1986computational}}: This is the most commonly used highly effective and complex compared to many other methods. It is a better method that without disturbing the features of the edges in the image afterwards and it apply the tendency to find edges and serious value of threshold. The algorithmic steps are as follows:
    \begin{enumerate}
        \item Convolve an image $f(r, c)$ with a Gaussian function to get smooth image $f(r, c). f(r, c)=f(r,c)*G(r,c,6)$.
        \item Apply first difference gradient operator to compute edge strength, then edge magnitude and direction are obtained as before.
        \item Apply non-maximal or critical suppression to the gradient magnitude. 
        \item Apply threshold to the non-maximal suppression image.
    \end{enumerate}
\end{enumerate}

\subsubsection{Corner Detection}
Corner detection works on the principle that if a small window is placed over an image, and that window is placed on a corner. If that window is moved in any direction, then there will be a large change in intensity.
\begin{enumerate}
    \item \textbf{The Moravec corner detection \cite{moravec1980obstacle}}: This is one of the earliest corner detection algorithms and defines a corner to be a point with low self-similarity. The algorithm tests each pixel in the image to see if a corner is present, by considering how similar a patch entered on the pixel is to nearby, largely overlapping patches.
    
    \item \textbf{Harris corner detector \cite{harris1988combined}}: This method was developed to identify the internal corners of an image and it is realized by calculating each pixel’s gradient. It is based on the local auto-correlation function of a signal which measures the local changes of the signal with patches shifted by a small amount in different directions.
    
    \item \textbf{Förstner corner detector \cite{forstner1987fast}}: This method solves for the point closest to all the tangent lines of the corner in a given window and is a least-square solution in order to achieve an approximate solution. The algorithm relies on the fact that for an ideal corner, tangent lines cross at a single point.
    
    \item \textbf{SUSAN (Smallest Uni-value Segment Assimilating Nucleus) corner detector \cite{smith1997susan}}: This method is based on brightness comparison and realized by a circular mask. If the brightness of each pixel within a mask is compared with the brightness of that mask's nucleus, then an area of the mask can be defined which has the same (or similar) brightness as the nucleus. The SUSAN area will reach a minimum while the nucleus lies on a corner point.
    
    \item \textbf{Fuzzy System \cite{banerjee2008handling}}: The measure of “cornerness” for each pixel in the image is computed by fuzzy rules (represented as templates) which are applied to a set of pixels belonging to a rectangular window. The possible uncertainty contained in the window-neighborhood is handled by using an appropriate rule base (template set). The fuzzy system is simple to implement and still fast in computation when compared to some existing fuzzy methods. Also, it can be easily extended to detect other features.
\end{enumerate}

\subsection{Quantum Methods in Edge Detection}
In the known classical edge extraction algorithms, for a typical image of $2^n$ x $2^n$ processing work with the computational complexity less than $O(2^{2n})$, it is impossible to complete such a task\cite{xu2020quantum}. By contrast, quantum image processing utilizes the superposition and entanglement characteristics of quantum mechanics to simultaneously calculate all the pixels of the image, thus realizing the acceleration of the algorithm. Basically, the quantum edge detection methods are the combination of QIR methods and classical edge methods. 

\begin{enumerate}
    \item \textbf{QSobel and other quantum image model based methods}: Early in 2014, Zhang et al. reported a novel quantum image edge extraction algorithm called QSobel \cite{article}. It is designed based on the FRQI and the famous edge extraction algorithm Sobel. Fan et al. proposed an image edge detection method based on Laplacian operator and Sobel in 2019 \cite{fan2019quantum}. In 2020, Xu et al. proposed a quantum image processing algorithm using edge extraction based on Kirsch operator \cite{xu2020quantum}. A quantum edge detection algorithm based on improved eight-direction Sobel operator was proposed by Ma et al. in 2020 \cite{ma2020demonstration}. 
    The flow chart is shown in the Figure \ref{fig:quantum_review}, which is divided into 4 stages more specifically. The classical digital image is quantized into quantum image NEQR model first. Then, the X-Shift and Y-Shift transformations are used to obtain the shifted image set. Following that, very pixel’s gradient is further calculated according to Sobel mask using the shifted image set simultaneously. Finally, the edge of the original image is extracted through the threshold operation $U_T$.
    
\begin{figure}[h!]
  \begin{center}
  \includegraphics[width=8cm]{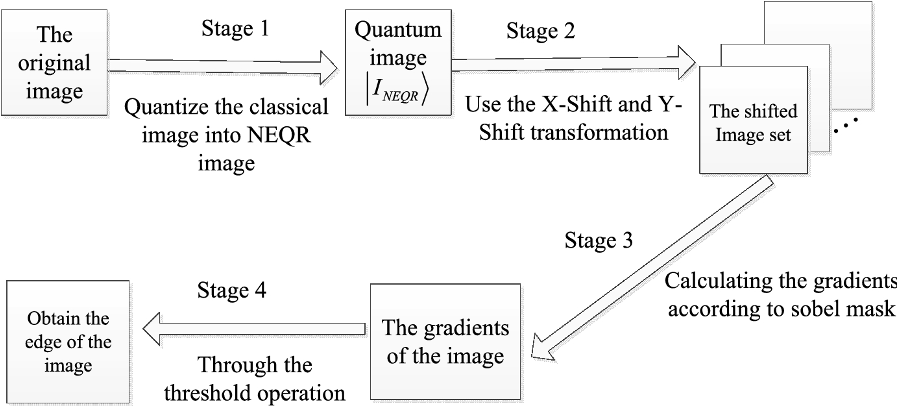}
  \end{center}
  \caption{Basic workflow of quantum image edge extraction, image from Fan et al.’s work\cite{fan2019quantum}}
  \label{fig:quantum_review}
\end{figure}

    \item \textbf{Quantum Hadamard Edge Detection (QHED)}: Another kind of quantum edge detection algorithm calculated the difference between adjacent pixels in the quantum image via Hadamard transform, this method is called Quantum Hadamard Edge Detection (QHED), which is reported by Yao et al. in 2017 \cite{2017}. This QHED algorithm encode the pixel values of the image in the probability amplitudes and the pixel positions in the computational basis states. In a N-pixel image, the pixels of the image are numbered in the form of $|b_1 b_2...b_{n-1} 0⟩$ ,where $b_i$ = 0 or 1. The positions of any pair of neighboring pixels in a picture column are given by the binary sequences $|b_1 b_2...b_{n-1} 0>$   and $|b_1 b_2...b_{n-1} 1>$, and the corresponding pixel intensity values are stored as the $c_{b_1 b_2...b_{n-1} 0}$   and $c_{b_1 b_2...b_{n-1} 1}$. By applying the Hadamard transform to the least significant bit on an arbitrary size quantum register. The total operation is then,

\begin{equation}
\mathrm{I}_{2^{\mathrm{n}-1}} \otimes \mathrm{H}=\frac{1}{\sqrt{2}}\left[\begin{array}{ccccccc}
1 & 1 & 0 & 0 & & 0 & 0 \\
1 & -1 & 0 & 0 & \ldots & 0 & 0 \\
0 & 0 & 1 & 1 & \cdots & 0 & 0 \\
0 & 0 & 1 &-1 & & 0 & 0 \\
& & \vdots & & \ddots & & \vdots \\
0 & 0 & 0 & 0 & \ldots & 1 & 1 \\
0 & 0 & 0 & 0 & & 1 & -1
\end{array}\right]
\end{equation}

where $I_{2^{n-1}}$ is the $2^{n-1}$ X $2^{n-1}$
unit matrix.

Applying this unitary to a quantum register containing pixel values encoded using the QPIE representation, as

\begin{equation}
\left(\mathrm{I}_{2^{\mathrm{n}-1}} \otimes \mathrm{H}\right) \cdot\left[\begin{array}{c}
\mathrm{c}_{0} \\
\mathrm{c}_{1} \\
\mathrm{c}_{2} \\
\mathrm{c}_{3} \\
\vdots \\
\mathrm{c}_{\mathrm{N}-2} \\
\mathrm{c}_{\mathrm{N}-1}
\end{array}\right] \rightarrow \frac{1}{\sqrt{2}}\left[\begin{array}{c}
\mathrm{c}_{0}+\mathrm{c}_{1} \\
\mathrm{c}_{0}-\mathrm{c}_{1} \\
\mathrm{c}_{2}+\mathrm{c}_{3} \\
\mathrm{c}_{2}-\mathrm{c}_{3} \\
\vdots \\
\mathrm{c}_{\mathrm{N}-2}+\mathrm{c}_{\mathrm{N}-1} \\
\mathrm{c}_{\mathrm{N}-2}-\mathrm{c}_{\mathrm{N}-1}
\end{array}\right].
\end{equation}

It is obvious that we now have access to the gradient between the pixel intensities of neighboring pixels in the form of $c_i - c_{i+1}$. The logic here is: When two pixels belong to the same region, their intensity values are identical and the difference vanishes, otherwise their difference is non-vanishing, which indicates a region boundary. 
This process results in the detection of horizontal boundaries between the even-pixels-pairs $0/1, 2/3$, and so on. For detection of horizontal boundaries between odd-pixel-pairs $1/2, 3/4$, etc., we can perform an amplitude permutation on the quantum register to convert the amplitude vector $(c_0,c_1,c_2,…,c_{N-1})^T$ to $(c_1,c_1,c_2,…,c_{N-1},c_0)^T$, and then applying the H-gate and measuring the quantum register conditioned on LSB being $|1\rangle$. 

\subsection{Image Classification}
\subsubsection{\textbf{Support Vector Machine(SVM) and Quantum SVM}}    
SVM classification is essentially a binary (two-class) classification technique and kernel methods can achieve more complexity in classical SVM. Similar to support vector machines, the Quantum SVM algorithm applies to classification problems that require a feature-map implicitly specified by a kernel. Quantum SVM using minimization quantum subroutine GroverOptim and provides convergence to a global optimum for non-convex cost function. Quantum SVM and kernels can exploit the higher dimensional space efficiently and can generate feature maps and subsequent decision boundaries that are difficult for classical kernel functions to match. As it mentioned in Park et al.'s work\cite{park2020practical}, Quantum SVM can potentially provide better performance if the underlying boundary is a complex one not captured well by traditional classical kernels.

\subsubsection{\textbf{K-means and Quantum K-means}} 
K-means algorithm belongs to the family of unsupervised machine learning algorithms that groups together n observations into K clusters making sure intra cluster variance is minimized. The most resource consuming operation for k-means algorithm is calculation of a distance between vectors. The quantum version of the K-means algorithm provides an exponential speed-up for very high dimensional input vectors by the fact that only log N qubits are required to load N-dimensional input vectors using the amplitude encoding\cite{kopczyk2018quantum}.

\subsubsection{\textbf{Convolutional Neural Network(CNN) and Quantum CNN}} 
CNN has shown very high performance in computer vision with the advantage of making good use of the correlation information of data. However, CNN is challenging to learn efficiently if the given dimension of data or model becomes too large. Cong extends the convolution layer and the pooling layer,the main features of CNN, to quantum systems as a model of Quantum CNN\cite{cong2019quantum}. The convolution circuit will find the hidden state by applying multiple qubit gates between adjacent qubits and the pooling circuit will reduce the size of the quantum system by observing the fraction of qubits or applying 2-qubit gates such as CNOT gates. After repeating the convolution circuit and pooling circuit until the size of the system is sufficiently small, the fully connected circuit can predict the classification result. The construction of the quantum convolutional layer can be divided into four steps: (1) The encoding process stores the pixel data corresponding to the filter size in qubits; (2) The learnable quantum circuits apply the filters that can find the hidden state from the input state; (3) The decoding process gets new classical data by measurement; (4) Repeat steps (1) to (3) to complete the new feature map. \cite{oh2020tutorial}
\end{enumerate}

\section{Datasets}
\subsection{MNIST Dataset}
This dataset \cite{lecun-mnisthandwrittendigit-2010} is an extensive set of $60000$ handwritten digits, every image has a  dimension of $28$x$28$ pixels. Along with training data it also provides $10000$ additional test dataset consisting of equally distributed digits from $0-9$.

\subsection{Fashion MNIST Dataset}
This dataset \cite{DBLP:journals/corr/abs-1708-07747} consists of $60000$ clothing images, $50000$ training data and $10000$ test dataset, with label from 10 different classes, every image has a  dimension of $28$x$28$ pixels. 
Here, for quantum binary classification, small dataset is used for training and subsequent testing of $3$ equally distributed clothing data class labelled as,
\begin{figure}[h!]
  \begin{center}
  \includegraphics[width=0.48\textwidth]{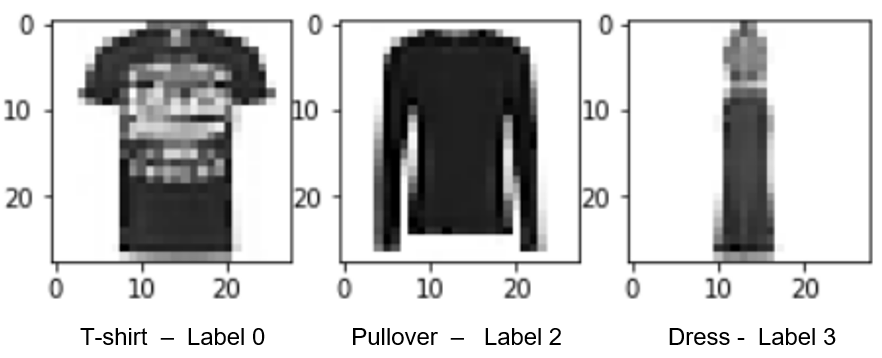}
  \end{center}
  \caption{Fashion MNIST data with label 0, 2, 3.}
  \label{fig:fashion MNIST}
\end{figure}

\subsection{CIFAR-10 Dataset}
This dataset (Learning Multiple Layers of Features from Tiny Images)\cite{cifar10} consists of $60000$ images distributed equally across $10$ different classes, each image is $32$x$32$ pixels out of which $10000$ images are used as test dataset. This dataset is divided in $6$ batches, $5$ batches are used for training the model and $1$ batch is used for model evaluation as a test dataset.

\subsection{Synthetic Datasets}
Current quantum devices (NISQ) have a limited number of qubits, thereby, limiting the exploration of images with large dimensions, to create synthetic datasets for evaluation of the quantum algorithms. Furthermore, we also plan to apply dimensionality reduction techniques such as Principal Component Analysis (PCA) on publicly available datasets before passing them through the quantum devices.

\section{Results}
\subsection{Classical Methods}

\subsubsection{Edge detection with MNIST dataset}

\begin{itemize}
    \item \textbf{Stage 1}: Normalization and Smoothing the image: An example of this stage is shown in Figure \ref{fig:stage-1}. 
    
    \item \textbf{Stage 2}: Getting edge related information using gaussian, laplace, sobel filter as demonstrated in Figure \ref{fig:stage-2}.
    
    \item \textbf{Stage 3}: Harris corner detector to detect edges in the digits as demonstrated in Figure \ref{fig:stage-3}.
    
    \item \textbf{Stage 4}: Conversion to bag-of-words representation.
    
    \item \textbf{Stage 5}: Histogram representation.
    
    \item \textbf{Stage 6}: Classification using SVM.
\end{itemize}

\begin{figure}[h!]
  \begin{center}
  \includegraphics[width=0.3\textwidth]{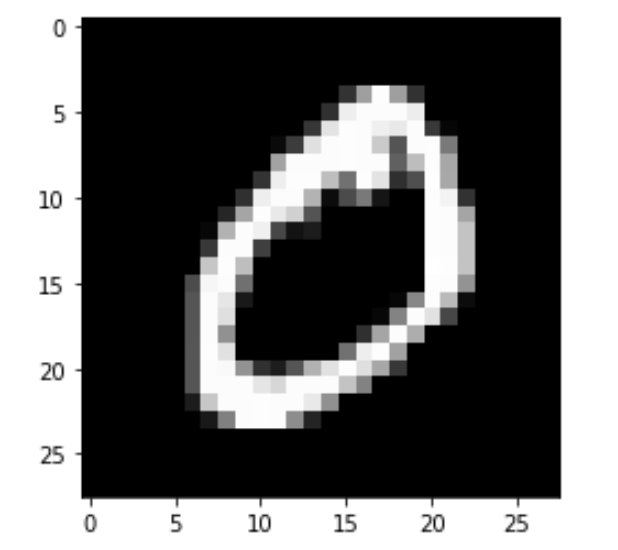}
  \end{center}
  \caption{stage-1 : Normalization and Smoothing the image}
  \label{fig:stage-1}
\end{figure}

\begin{figure}[h!]
  \begin{center}
  \includegraphics[width=0.5\textwidth]{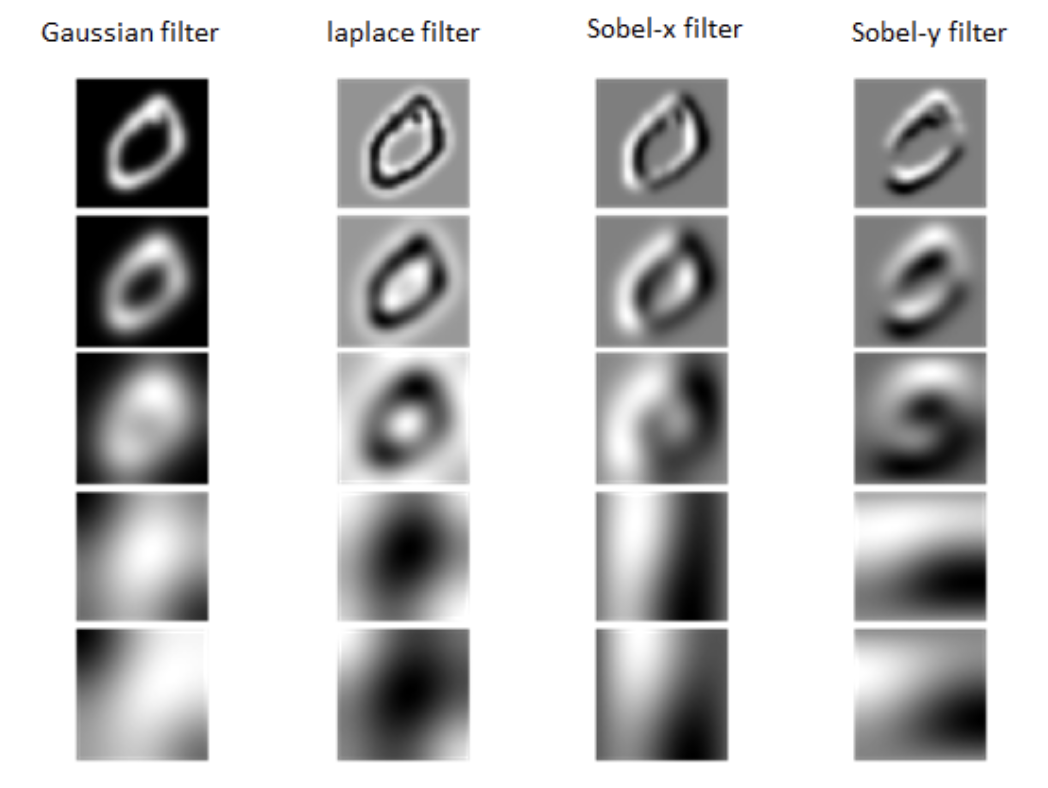}
  \end{center}
  \caption{stage-2 : Getting edge related information using gaussian, laplace, sobel filter.}
  \label{fig:stage-2}
\end{figure}

\begin{figure}[h!]
  \begin{center}
  \includegraphics[width=0.3\textwidth]{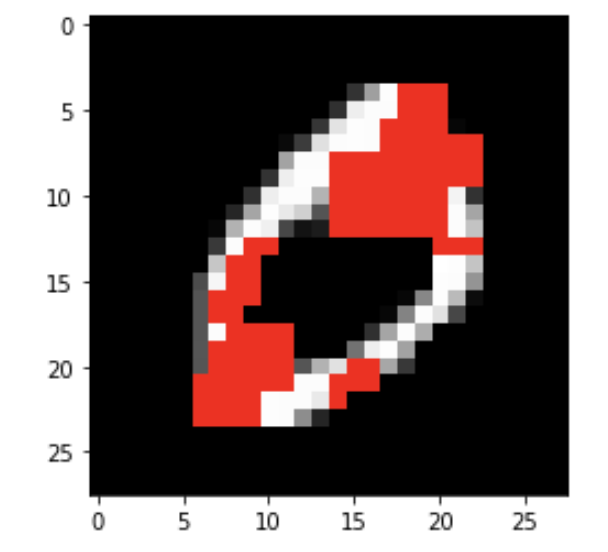}
  \end{center}
  \caption{stage-3 : Harris corner detector to detect edges in the digits.}
  \label{fig:stage-3}
\end{figure}

\begin{figure}[h!]
  \begin{center}
  \includegraphics[width=0.5\textwidth]{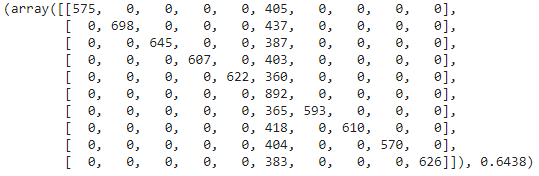}
  \end{center}
  \caption{Confusion Matrix and Prediction accuracy on Test dataset of MNIST}
  \label{fig:stage-3}
\end{figure}

\subsubsection{Deep Learning Model for MNIST dataset}
We use the architecture with Resnet \cite{he2015deep} blocks, Batch normalization and ReLU activation for classification of MNIST dataset. The architecture provides an accuracy of 99.6\% in only 12 epochs. We used the Learning Rate Finder provided by FastAI library \cite{fastai2020} and found that the best learning rate for our architecture is 0.05. We used Stochastic Gradient Descent (SGD) as the optimizer and the loss function as Cross Entropy Loss as MNIST is a multi-class classification problem. 

\subsubsection{Deep Learning Model for CIFAR-10 dataset}
Here we have used SOFTMAX activation function, with 10 units in output layers. In compiling of model, we took account for loss function using Sparse Categorical-Cross Entropy as classes are totally distinctive. We have used Adam Optimizer to adapt learning rate, as it updates weights after every iteration. After running 15 epochs, final results obtained with accuracy of 77.5\%. 

\begin{figure}[h!]
  \begin{center}
  \includegraphics[width=0.5\textwidth]{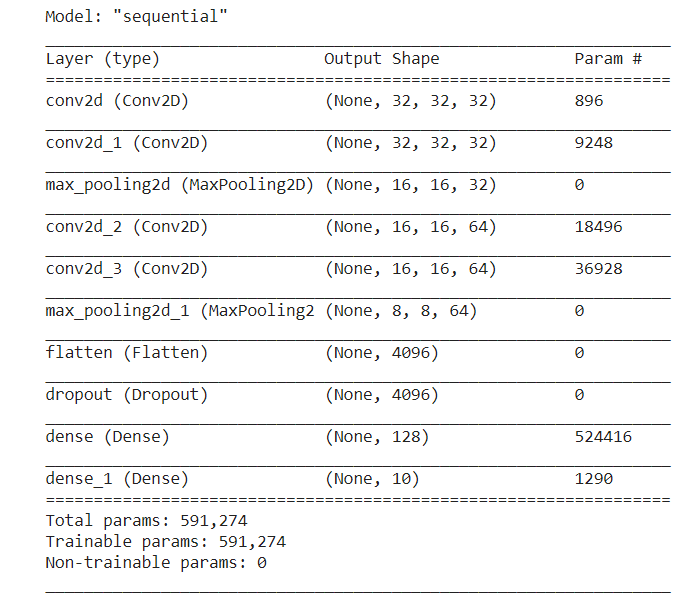}
  \end{center}
  \caption{Deep learning model summary (CNN)}
  \label{fig:Cifar10_model}
\end{figure}

\begin{figure}[h!]
  \begin{center}
  \includegraphics[width=0.5\textwidth]{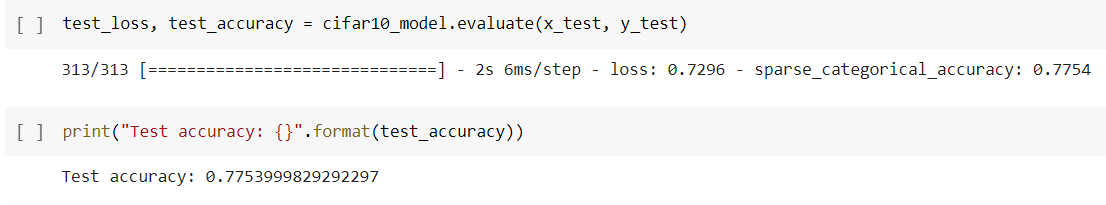}
  \end{center}
  \caption{Deep learning model results (CNN)}
  \label{fig:Cifar10_results}
\end{figure}

\subsection{Quantum Methods}

\subsubsection{ Quantum Image binary classification of MNIST dataset}
Here, IBM Quantum lab platform is used for binary classification of fashion MNIST image data.\cite{QHEDblock2020}
In the method, at first classical data was split into sample, validation and test data and labels, followed by preprocessing classical dataset with feature dimension 5 using standardization, principal component analysis and normalization. 
For multi-class classification we have used three quantum support vector machines based binary classifiers with One-vs-Rest approach classifying a particular label e.g., label 0 as positive (1) and rest as negative (0).
In second stage, classical data was encoded to quantum state using different quantum feature Map, followed by preparing quantum kernel matrices utilising quantum state-vector simulator for training and validation dataset. Finally, accuracy of quantum kernel matrices was estimated on validation and test dataset with prediction probability of particular label for test data. 

\begin{figure}[h!]
  \begin{center}
  \includegraphics[width=0.5\textwidth]{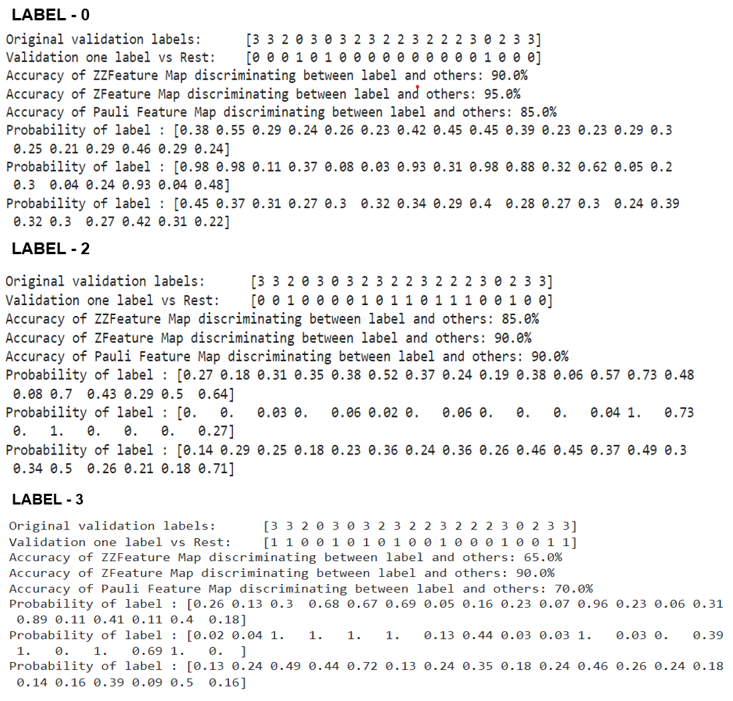}
  \end{center}
  \caption{Prediction probability of Image label based on different feature map}
  \label{fig:Prediction probability of Image label based on different feature map}
\end{figure}
Finally, the result was obtained using the different quantum feature map based quantum kernels in correctly identifying and predicting label in the test data based on the highest probability corresponding to a particular label.
The Quantum binary image classification was performed based on different feature map i.e. ZZFeature, Pauli and ZFeature map results were compared for the obtained accuracy of classifying different image labels.

\begin{figure}[h!]
  \begin{center}
  \includegraphics[width=0.5\textwidth]{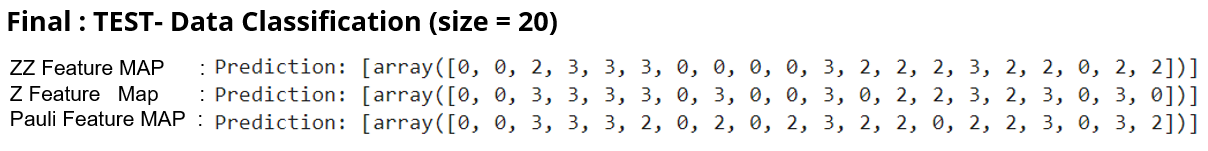}
  \end{center}
  \caption{Quantum Binary Image classification: Test result Label 0, 2 ,3  }
  \label{fig:Quantum Binary Image classification: Test result Label 0, 2 ,3}
\end{figure}

\begin{figure}[h!]
  \begin{center}
  \includegraphics[width=0.48\textwidth]{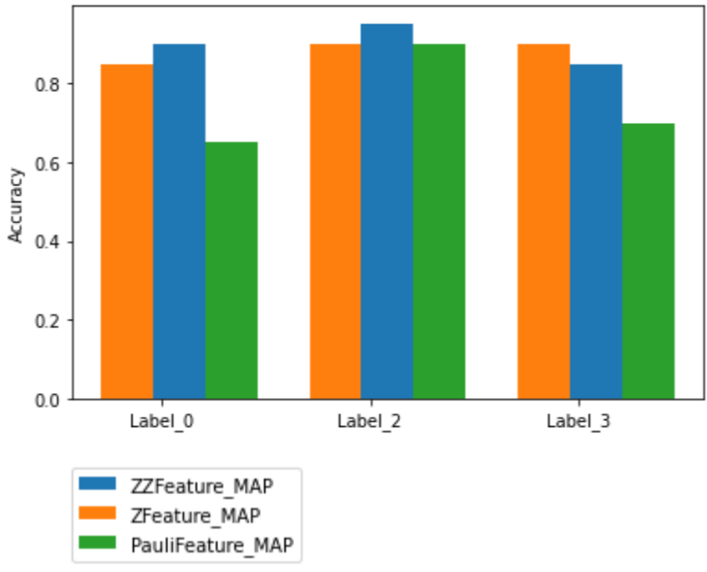}
  \end{center}
  \caption{Different feature map based quantum kernel accuracy in prediction}
  \label{fig:Different feature map based quantum kernel accuracy in prediction}
\end{figure}

Quantum kernel based on ZZfeature map involves entangling blocks in different configuration i.e linear, circular and full entanglement for encoding classical data. This comparison shows prediction accuracy within ZZfeature map with different entanglement configuration of test image data labels, which helped in realization of achieving suitable configuration for 3-class image classification with high accuracy.

\begin{figure}[h!]
  \begin{center}
  \includegraphics[width=0.5\textwidth]{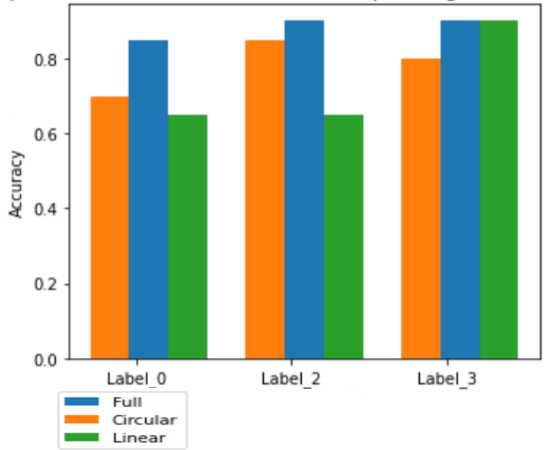}
  \end{center}
  \caption{Prediction based on different entanglement configuration of ZZFeature map}
  \label{fig:Prediction based on different entanglement configuration of  ZZFeature map}
\end{figure}

\subsubsection{Quantum Image Classification using Edge detection }
This method is split into two stages, the first stage includes finding edges of the image and in the second stage the edges are then used for image classification. The edge detection with classical algorithms is much slower compared to hadamard based edge detection.  The time complexity of classical operation is O($2^n$) however the quantum system can perform the gradient calculation between pixels much faster using hadamard operation O($mn.log(mn)$). Hence leveraging the faster operating speed this method performed edge detection using quantum circuits and performed classification using Logistic Regression was used to get the accuracy of the model. For image representation, QPIE (Quantum Probability Image Encoding) is used, where the image is using n = $log_2 N$ qbits to represent an image where N is the number of pixels from the image.

Results:

\begin{figure}[h!]
  \begin{center}
  \includegraphics[width=0.5\textwidth]{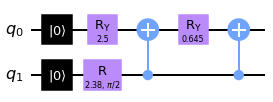}
  \end{center}
  \caption{QPIE image representation}
  \label{fig:MNIST_ZZfeatMap_CKT}
\end{figure}
\begin{figure}[h!]
  \begin{center}
  \includegraphics[width=0.5\textwidth]{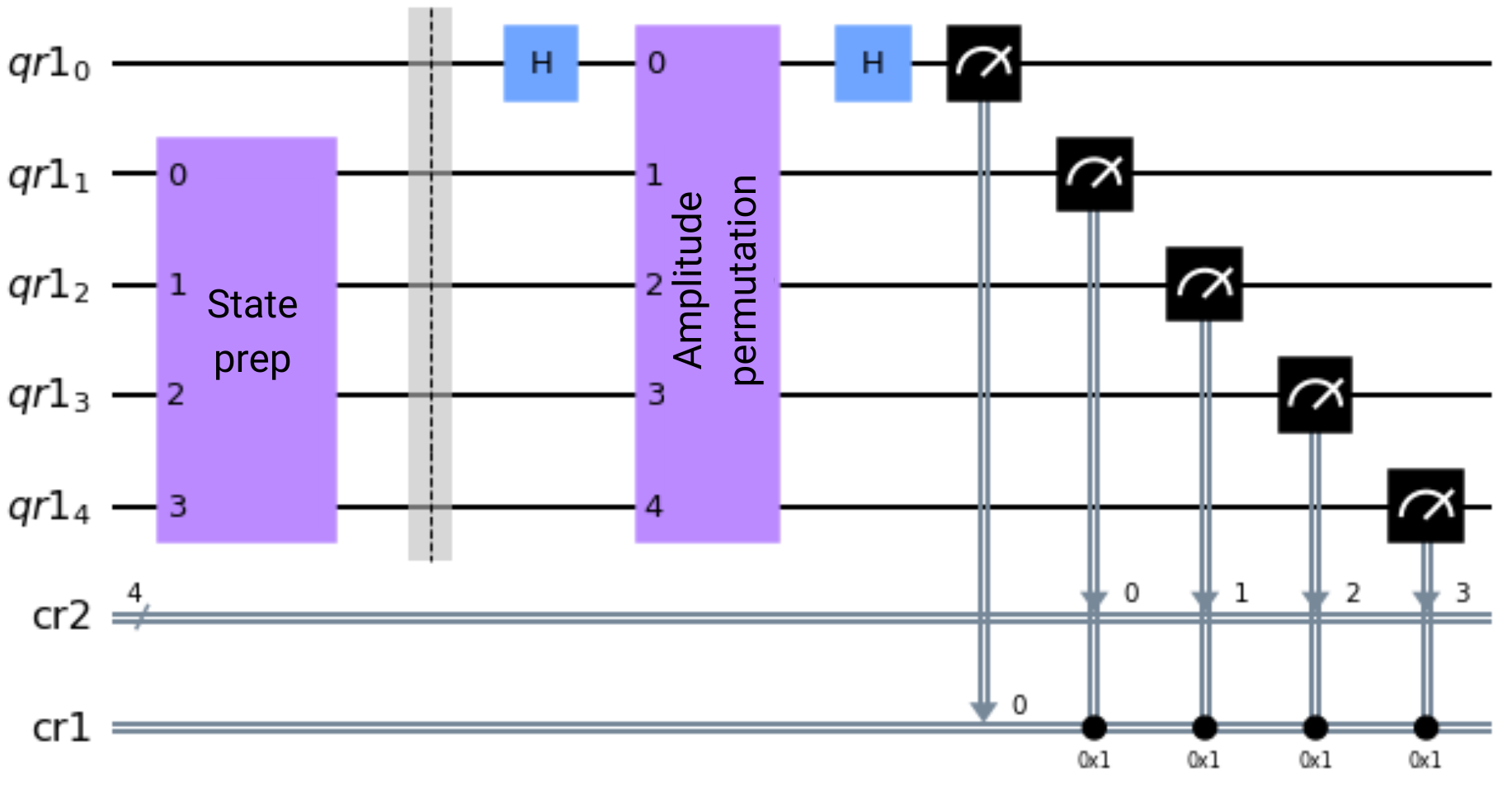}
  \end{center}
  \caption{Quantum Hadamard Edge detection using Simulator}
  \label{fig:MNIST_ZZfeatMap_CKT}
\end{figure}
The Algorithm performed two scan for gradient calculations, in first scan only horizontal edges were detected and in second scan the vertical edges were detected. This operation is equivalent to SobelX and SobelY operation in classical algorithms.
\begin{figure}[h!]
  \begin{center}
  \includegraphics[width=0.5\textwidth]{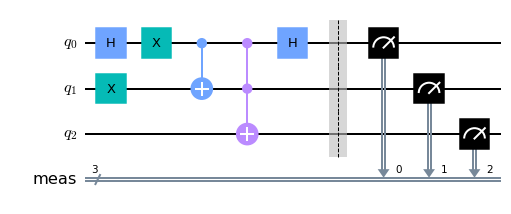}
  \end{center}
  \caption{Quantum Hadamard Edge detection horizontal scan}
  \label{fig:MNIST_ZZfeatMap_CKT}
\end{figure}
\begin{figure}[h!]
  \begin{center}
  \includegraphics[width=0.5\textwidth]{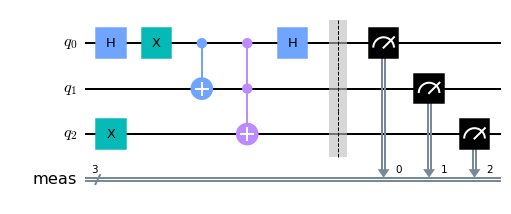}
  \end{center}
  \caption{Quantum Hadamard Edge detection Vertical scan}
  \label{fig:MNIST_ZZfeatMap_CKT}
\end{figure}
The method, also handles large images by splitting them in smaller patches and applies edge detection separately. This is majorly done due to limitation of qbits in actual quantum computers, coupling between qbits and also to reduce the probability of noise impacting the operation. For MNIST dataset, the image is scaled to 32x32 pixels and then divided into 4x4 pixel patches, the reason for choosing 32 is to provide support for CIFAR dataset and also increase the quality of edge detection because of up scaling.
\begin{figure}[h!]
  \begin{center}
  \includegraphics[width=0.5\textwidth]{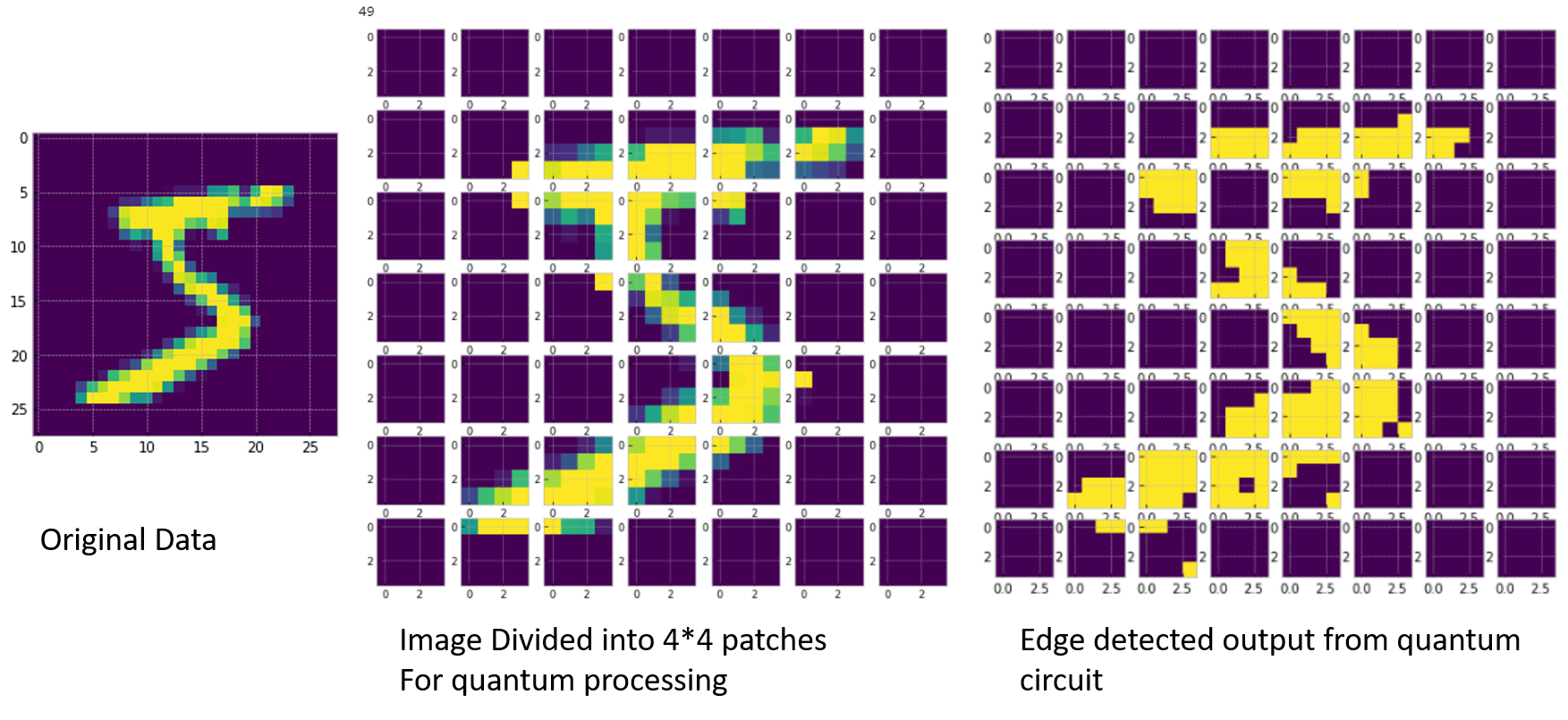}
  \end{center}
  \caption{The flow of QHED with large images.}
  \label{fig:MNIST_ZZfeatMap_CKT}
\end{figure}
Coming to stage 2 of the method, classification is done using a logistic regression model with mutli-class model including intercept term with Lasso regression as penalty. 
\begin{figure}[h!]
  \begin{center}
  \includegraphics[width=0.5\textwidth]{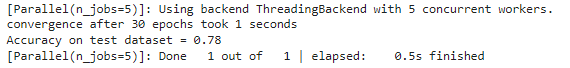}
  \end{center}
  \caption{Accuracy of Method2 with sample MNIST dataset}
  \label{fig:MNIST_ZZfeatMap_CKT}
\end{figure}
The Cifar dataset is more complex compared to the MNIST dataset hence requires more time for edge detection, To perform training the scope was reduced by using 100 samples to train from training dataset and the testing was performed using 10 samples of test dataset. which resulted in 10\% accuracy. The reduce in accuracy can be linked to the incorrect splitting ratio of the image to small samples which was used to detect the edges and also due to smaller samples of training. 
\begin{figure}[h!]
  \begin{center}
  \includegraphics[width=0.5\textwidth]{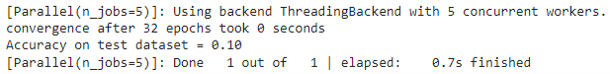}
  \end{center}
  \caption{Accuracy of Method2 with sample CIFAR dataset}
  \label{fig:MNIST_ZZfeatMap_CKT}
\end{figure}
\subsubsection{Impact of Noise}
In the Noisy Intermediate-Scale Quantum (NISQ) era, noise in quantum computers plays a crucial role. We study the impact of various noise models for the Quantum Edge Detection experiment on MNIST dataset.

\begin{itemize}
    \item \textbf{Depolarizing Noise}: The depolarizing channel is defined as the following:
    \begin{equation}
        E(\rho)=(1-\lambda) \rho+\lambda \operatorname{Tr}[\rho] \frac{I}{2^{n}}
    \end{equation}
    with $0 \leq \lambda \leq 4^{n} /\left(4^{n}-1\right)$, where  $\lambda$ is the depolarizing error parameter and $n$ refers to the number of qubits. Depolarizing Noise creates spontaneous transitions among eigenstates, inducing the sudden death of maximally entangled states \cite{bavontaweepanya2018effect}. We analyze the impact of depolarizing noise on single qubit gates, two-qubit gates and the combined effect on both types of gates. Figure \ref{fig:D_NOISE_1} and \ref{fig:D_NOISE_2} demonstrate the impact of depolarizing noise with various noise probabilities. We notice that noise on both the gates combined has a larger impact on the performance of the model.

\begin{figure}[h!]
  \begin{center}
  \includegraphics[width=0.5\textwidth]{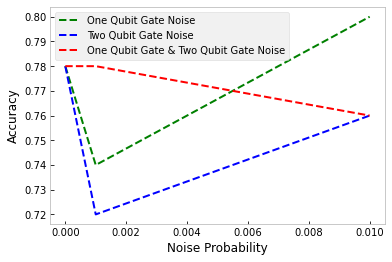}
  \end{center}
  \caption{Impact of Depolarizing Noise with a Noise Probability of 0.001, 0.01}
  \label{fig:D_NOISE_1}
\end{figure}
\begin{figure}[h!]
  \begin{center}
  \includegraphics[width=0.5\textwidth]{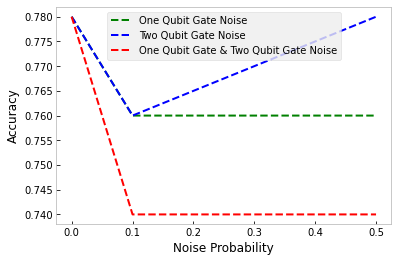}
  \end{center}
  \caption{Impact of Depolarizing Noise with a Noise Probability of 0.1, 0.5}
  \label{fig:D_NOISE_2}
\end{figure}

    \item \textbf{Pauli Noise}: The amplitude and the phase of the qubit flip based on the noise parameter provided to the Pauli Noise Model. We analyze the impact of bit flip noise, phase flip noise and the combined effect of both the noises on the Quantum Edge Detection for MNIST dataset. Figure \ref{fig:P_NOISE_1} demonstrate the performance of the model with various Pauli Noise probabilities. We notice that noise probability 0.001 has a large impact on the model's performance, decreasing the accuracy from 0.74 to 0.68. 
    
    \begin{figure}[h!]
  \begin{center}
  \includegraphics[width=0.5\textwidth]{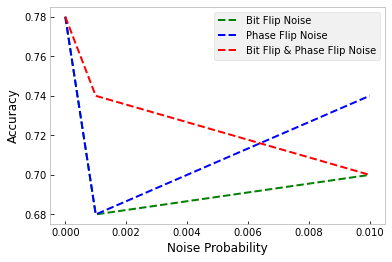}
  \end{center}
  \caption{Impact of Pauli Noise with a Noise Probability of 0.001, 0.01}
  \label{fig:P_NOISE_1}
\end{figure}

    Figure \ref{fig:NOISE_COMP} compares the edge detection results for model with no noise, depolarizing noise and pauli noise. We notice that after adding different noises, the digits are still recognizable. Figure \ref{fig:NOISE_COMP_2} demonstrates the performance of the edge detector classifier with training data from depolarizing noise model and pauli noise model. Pauli Noise has a worse impact on the model's accuracy.

 \begin{figure}[h!]
  \begin{center}
  \includegraphics[width=0.5\textwidth]{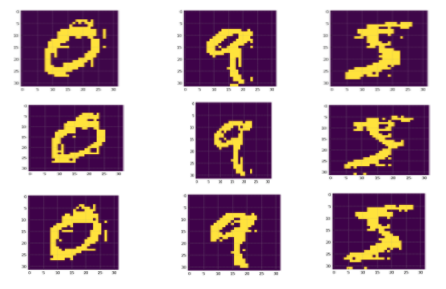}
  \end{center}
  \caption{Comparing the edge detection results for Noiseless model (top 3 images), Depolarizing Noise model with Noise Probability 0.5 for single qubit and two-qubit gates (middle 3 images), Pauli Noise model with Noise Probability 0.01 for both bit flip and phase flip (bottom 3 images)}
  \label{fig:NOISE_COMP}
\end{figure}

\begin{figure}[h!]
  \begin{center}
  \includegraphics[width=0.52\textwidth]{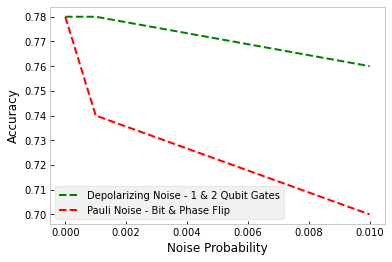}
  \end{center}
  \caption{Comparing the model performance with depolarizing noise and pauli noise}
  \label{fig:NOISE_COMP_2}
\end{figure}

\end{itemize}

\section{Conclusion}
In quantum image processing, we have used different image encoding methods for edge detection and binary image classification. When compared to classical process, quantum provides advantage in runtime, space complexity i.e. number of bits/qubits required for image encoding, but is not sufficient in terms of depth and width of image. With Quantum methods it is quite challenging to handle larger images as circuit design gets complex with increased number of qubits required, leading to addition of noise with inaccurate edge detection and classification results. In this project, we have used gate based quantum statevector and qasm simulator that are slower in performance which in-turn increases the training time for larger datasets and requires splitting of data into smaller sets and recombining to get higher accuracy for larger image pixel and datasets. While running real quantum may provides better performance but it is more exposed to the noise from the devices.
As advancements with this projects, our team is working to analyse impact of different noise models on the quantum image encoding circuit and its results when processing smaller pixel as well as large pixel images. Our work in future will include implementing from quantum processed results for image verification with original/real-time images to reach better accuracy.

\bibliography{references}
\bibliographystyle{unsrt}

\end{document}